# Importance of nuclear triaxiality for electromagnetic strength, level density and neutron capture cross sections in heavy nuclei

E. Grosse [b], A.R. Junghans [a] and R. Massarczyk [a,b]
[a] Institute of Radiation Physics, Helmholtz-Zentrum Dresden-Rossendorf, 01328 Dresden, Germany
[b] Institute of Nuclear and Particle Physics, TU Dresden, 01069 Dresden, Germany

**Abstract:**
Cross sections for neutron capture in the range of unresolved resonances are predicted simultaneously to average level distances at the neutron threshold for more than 100 spin-0 target nuclei with A >70. Assuming triaxiality in nearly all these nuclei a combined parameterization for both, level density and photon strength is presented. The strength functions used are based on a global fit to IVGDR shapes by the sum of three Lorentzians adding up to the TRK sum rule and theory-based predictions for the A-dependence of pole energies and spreading widths. For the small spins reached by capture level densities are well described by only one free global parameter; a significant collective enhancement due to the deviation from axial symmetry is observed. Reliable predictions for compound nuclear reactions also outside the valley of stability as expected from the derived global parameterization are important for nuclear astrophysics and for the transmutation of nuclear waste.

## 1 Introduction

The radiative capture of neutrons in the keV to MeV range by heavy nuclei plays an important role in considerations for advanced systems aiming for the reduction of radioactive nuclear waste [1]. This process is of interest also for the cosmic nucleosynthesis, especially for scenarios with high neutron fluxes, where neutron capture processes lead to a production of nuclides beyond Fe [2]. Predictions for radiative neutron capture cross sections in the range of unresolved resonances are based on statistical model calculations. Their reliability depends not only on the proper characterization of the input channel, but more strongly on the details determining the decay of the intermediately formed compound nucleus. Here the strength of its electromagnetic decay is of importance as well as the open phase space in the final nucleus, i.e. the density of levels reached by the first photon emitted. The experimental studies forming the basis for parameterizations can mainly be performed on nuclei in or close to the valley of beta-stability, but in cosmic environments many radiative processes occur in exotic nuclei which are not easily accessible experimentally. Similarly the knowledge of radiative neutron capture cross sections in actinide and other unstable nuclei is of importance for the understanding of the competition between nuclear fission and the production of long-lived radionuclides by capture. It is thus desirable to derive a parameterization which is global as based on concepts accepted to be valid generally and thus expected to be applicable also away from the stable nuclei. As is well known [3], the variation of nuclear quadrupole moments over the nuclide chart is very significant. Thus it is justified to investigate the influence of nuclear shapes on the extraction of strength functions from isovector giant dipole resonance (IVGDR) data as well as on nuclear level densities. If the restriction to axial symmetry is released, the contribution of collective rotation to level densities is significantly increased [3, 4], and Lorentzian fits to IVGDR data are improved [5].

Previously the results of various experiments on electromagnetic processes were often analysed [3] not regarding triaxiality. As demonstrated [6] one has to go considerably beyond the well documented [7] information on B(E2)-values and their relation *e.g.* to quadrupole moments. Also theoretically the breaking of axial symmetry has often been disregarded, although it was shown [8] within the Hartree-Fock-Bogoliubov (HFB) scheme, that exact 3-dimensional angular momentum projection results in a pronounced triaxial minimum also for deformed nuclei. Various spectroscopic studies (*e.g.* [6, 9, 10, 11, 12]) have identified triaxiality effects in very many nuclei and especially in nuclei with small but non-zero quadrupole moments. The study presented in the following makes use of a constrained CHFB-calculation for more than 1700 nuclei [13], which predicts not only quadrupole transitions rather well, but also the breaking of axial symmetry, *i.e.* the triaxiality parameter $\gamma$. Predictions derived using these results in the parameterization for the energy dependence of photon strengths as well as of nuclear level densities will be compared to average radiative widths at the neutron separation energy $S_n$ and of capture cross sections in the energy range of 30 keV. The present investigation tests a global prediction for 132 nuclides reached by neutron capture in spin-0 targets.

## 2    Dipole strength in triaxial nuclei

Electromagnetic processes play an important role not only in nuclear spectroscopy but also for the de-excitation processes following neutron capture or other nuclear reactions. Since decades the relation of the IVGDR to the nuclear radiative (or photon) strength [14, 15] is considered the basis of its parameterization for heavy nuclei. Its mean position $E_0$ can be predicted using information from liquid drop fits to ground state masses [16] and for triaxial nuclei the three pole energies $E_k$ are given by *a priori* information on the three axis lengths $r_k$ : $E_k = r_0/r_k \cdot E_0$. A parameterization of the electromagnetic strength in heavy nuclei with mass number A>70, which considers their triaxial deformation, was shown to be in reasonable accordance to various data of photon strengths $f_1(E_\gamma)$ [5]. This triple Lorentzian (TLO) approach [18, 19], combined to the axis ratios calculated by CHFB [13], describes the shapes of their IVGDR's as well as their low energy tail at energies below the neutron separation energy $S_n$. Using averages from the even neighbours this is the case also for odd nuclei as reached by capture from even target nuclei and Eq. (1) describes both cases (with the fine structure constant $\alpha$):

$$f_{E1}(E_\gamma) = \frac{4\alpha}{3\pi\, g_{eff}\, m_N c^2} \frac{ZN}{A} \sum_{k=1}^{3} \frac{E_\gamma \Gamma_k}{(E_k^2 - E_\gamma^2)^2 + E_\gamma^2 \Gamma_k^2}; \qquad g_{eff} = \sum_{r=1}^{2\cdot\min(\lambda,J_0)+1} \frac{2J_r+1}{2J_0+1} = 2\lambda+1 = 3. \qquad (1)$$

To fix its low energy tail of importance for radiative capture processes only its widths $\Gamma_k$ have to be known in addition to its full strength – fixed by the TRK sum rule for the nuclear dipole ($\lambda=1$) strength [18- 21]. Here the relation between GDR pole energy and width, well-established by hydrodynamics, can be generalized for triaxial shapes [22]: $\Gamma_k = c_w \cdot E_k^{1.6}$ with the proportionality factor $c_w \cong 0.45$ resulting from a fit to data for many nuclei with 70<A<240. For two nuclei often considered spherical the TLO sum for the IVGDR is compared in Fig. 1 to rescaled [24] data; the three poles are indicated as black bars. Obviously the fit is in accord to the prediction – in contrast to previous Lorentzian fits [23, 26], which clearly exceed the TRK sum rule, and their difference to TLO increases with decreasing photon energy [5]. This feature is of large importance for radiative capture which populates an excitation energy region of high level density $\rho(E_x)$, when $E_x$ is close to $S_n$, i.e. $E_\gamma$ is small. At such small energy $f_1(E_\gamma)$ is determined for TLO solely by the width parameter and the axis ratios from CHFB are not essential, but support the validity of the TRK sum rule. When account is made for instantaneous shape sampling (ISS) [24] due to the variance of the deformation parameters [13] TLO describes the IVGDR peak even better. In nearly all cases studied so far the TLO prediction is below experimental data [19, 24] acquired by photon scattering or other radiative processes under adoption of the Axel-Brink hypothesis [15, 25]. Thus clear experimental evidence is missing which would imply a need for energy dependent strength reductions proposed on the basis of IVGDR fits neglecting triaxiality [23, 26].

At gamma-energies below the neutron binding energy $S_n$ photon strength components, which are not of isovector electric dipole character, contribute to radiative capture [23, 26-30]. Respective information from photon scattering [31-33] is of use, asserting equal integrated strength for collective modes based on nuclear ground states and those on top of excited states [15, 25]. Minor strength, partly of other multipolarity, may also be derived from the analysis of gamma-decay following nuclear reactions [34-36] and our analysis investigates its importance. Two such components, both depending on the deformation β, have considerable impact on the predictions for radiative capture, as shown in in Fig. 1 and later in Ch. 4:

1. Orbital magnetic dipole strength (scissors mode [32, 36]), which is approximated to peak at $E_{sc} = 0.21 \cdot E_0$ with a maximum strength of $Z^2 \cdot \beta^2/45$ GeV$^{-3}$.
2. Electric dipole strength originating from coupled $2^+$ and $3^-$-phonons [31] is assumed to peak around $E_{quad} + E_{oct} = E_{qo} \approx$ 2-4 MeV with a height of $Z \cdot A \cdot \beta^2 \cdot E_{qo}/200$ GeV$^{-3}$.

For both a Gaussian distribution with $\sigma$ = 0.6 MeV is assumed and it is admitted, that the guesses as presented here can only serve as a very first hint on the eventual role of these strength components.

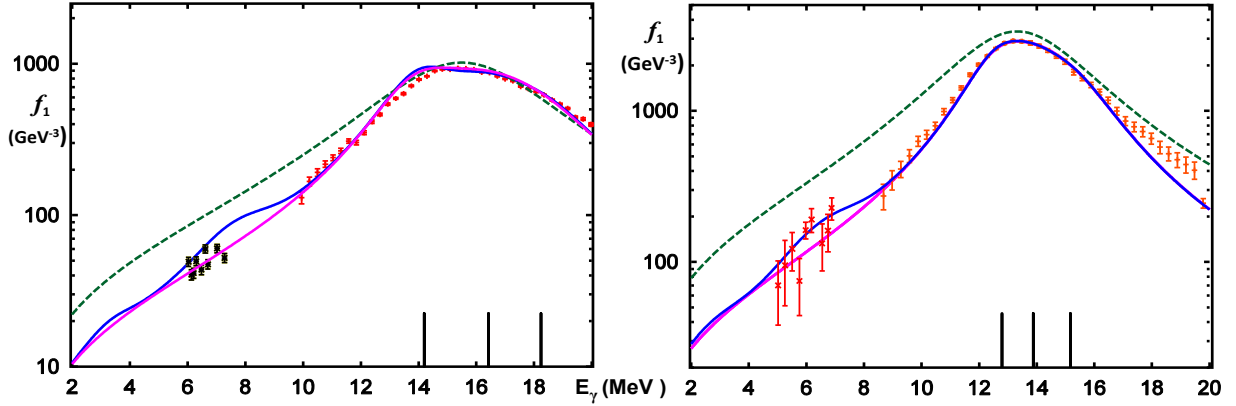

**Fig. 1:** The photon strength in comparison to a SLO-fit (dashed green) and TLO (magenta) with ISS, which is not included in the lines depicting the sum of 'minor' components and TLO (blue).
**Left panel**: The data above $S_n$ are from photo-neutron data on $^{nat}$Ag [34] and the ones below are derived from gamma decay subsequent to resonant neutron capture $^{105}$Pd(n,$\gamma$) [26].
**Right panel:** Photon strength derived from the photo-neutron cross section (+, [34]) on $^{197}$Au; also shown are photon scattering data (×, [15]) obtained with a quasi-monochromatic beam.

## 3  Level densities in triaxial nuclei

Since long the experimentally observable level density $\rho(E_x)$ is known to change strongly with nuclear deformation: An enhancement of $\rho(E_x)$ caused by allowing rotational bands on top of each intrinsic state was predicted [3, 4] to depend on shape asymmetry and in the limit of low spin $I$ one gets:

$$\rho(E_x,I) \longrightarrow \frac{2I+1}{\sqrt{8\pi}\,\sigma}\omega(E_x) \text{ for axial; } \rho(E_x,I) \longrightarrow \frac{2I+1}{4}\omega(E_x) \text{ for triaxial shapes, assuming } \Re\text{-symmetry} \quad (2).$$

Compared to the spherical case the enhancement is around 50 for one rotational degree of freedom (axial case) and this is considered 'standard' enhancement [23]. But, as obvious from Eq. (2), the effect of two extra rotational axes amounts to another factor of $\approx$ 6, when a typical spin dispersion (or cut off) factor of $\sigma \approx 5$ is assumed. Surprisingly such a large collective enhancement has not yet been included in comparisons to respective data, and a seemingly satisfying agreement was reached without by extra means. But the novel finding of triaxiality being a very widespread property of heavy nuclei [5, 9-13] calls for a compensating reduction in the prediction for the intrinsic state density $\omega(E_x)$.

It was proposed [17, 23, 37] to distinguish between a superfluid (quasi-Bosonic) phase below and a Fermi gas description above a transition temperature $t_t = \Delta_0 \cdot e^C/\pi = 0.567 \cdot \Delta_0$, with the Euler constant C and the paring gap given by $\Delta_0 = 12 \cdot A^{-1/2}$. In both regimes the intrinsic state density $\omega(E_x)$ is related to the nuclear entropy $S$ with an additional term containing the determinant $d$ of the matrix resulting from the use of the saddle point approximation [3, 23, 37]:

$$\omega(E_x) \approx \frac{e^S}{\sqrt{d}} \quad (3).$$

Sufficiently above $t_t$ the entropy $S$ is proportional by 2a to the temperature parameter $t$. The "level density parameter a" has a main component given by the mass number $A$ divided by the Fermi energy $\varepsilon_F = 37\ MeV$, in correspondence to the expectation for nuclear matter [3]:

$$a = \frac{\pi^2 A}{4\ \varepsilon_F} + \frac{A^{2/3}}{11} \qquad (4)$$

It is enlarged by a small surface term, which in our approach is the only free parameter used to arrive at an average agreement to neutron capture resonance data [23]. The widespread habit to further modify a – proposed as phenomenological inclusion of shell effects or even taken as a free local fit parameter [17, 23, 37] – is avoided here to suppress any mutual interference between the A and $E$-dependence of $\omega(E_x)$. The energy shift related to pairing is A-dependent and is usually [23] quantified by pairing gap $\Delta_0$ and condensation energy $E_{con} = \frac{3a}{2\pi^2}\Delta_0^2$ [17, 23, 37], often reduced by an additional shift δ [23, 37]. As shown in Eq. 6 the back-shift $E_{bs}$ we take is the difference of shell correction and $E_{con}$, and the zero energy for the Fermi gas is shifted by $E_{bs}$ from the excitation energy $E_x$. This ansatz differs from shifts used previously [23, 37], but it avoids the inconsistencies in the description of pairing effects, which appear for light nuclei in earlier work – as recently demonstrated [39]. Here the reduction resulting from the large shift counteracts the enhancement in level density due to triaxiality.

Shell effects and the odd-even mass difference are accounted for using the shell correction $\delta W_o$ as compiled for RIPL-3 [23] and taken from the mass calculation performed with the Myers-Swiatecki formalism [40]; in the table of ref. [23] the deformation energy calculated within the liquid drop model is also given and it is subtracted here to account for ground state deformation. The shell correction is reduced with increasing temperature parameter t (*i.e.* excitation) as shown in eq. 5. This procedure is at variance to previous work [23], but similar as discussed [3] and applied before [41].

$$\delta W(t) = \delta W_0 \cdot \frac{\tau^2 \cosh \tau}{(\sinh \tau)^2} \xrightarrow[t \to 0]{t \to \infty} 0 \atop \delta W_0 \ ; \quad \tau = \frac{2\pi^2 t}{\hbar \varpi_{sh}} A^{1/3} ; \quad \hbar \varpi_{sh} \cong \frac{1.4 \hbar^2}{r_0^2 m_N A^{1/3}} \qquad (5).$$

In the Fermi gas regime ($t > t_t$) one gets for entropy $S$, energy $E_x$ and determinant $d$:

$$S = 2at - \frac{\delta W(t)}{t} + \frac{\delta W_0}{t}\frac{\tau}{\sinh(\tau)} \xrightarrow[t \to 0]{t \to \infty} 2at \atop 0 \ ; \quad E_x = at^2 - \delta W(t) + E_{con} \xrightarrow[t \to 0]{t \to \infty} E_{con} + at^2 \atop E_{con} - \delta W_0 \ ; \quad d = \frac{144}{\pi} a^3 t^5 \qquad (6).$$

As obvious, the damping does not depend on any additional parameter as it is determined by the average frequency $\varpi_{sh}$ of the harmonic oscillator determined by radius parameter $r_0$ and nucleon mass $m_N$ alone. Additionally the limits for large and small $t$ are determined separately for $S$ and $E_x$ (cf. Eq. (6)) and are thus under independent control. The reduction of the number of free parameters to the one in Eq. (4) is a clear advantage over previous proposals for analytic level density models [23, 37].

Knowing $\delta W_0$ the intrinsic state density $\omega(E_x)$ can be calculated from Eqs. (3) to (6) for the Fermi gas regime as well as the values for $S$, $E_x$ and $d$ at the point of transition. Below $E_t = E_x(t_t)$ an interpolation to the ground state ($E_x$=0 MeV, $S = S_0$) is used and with $S_0$=n· ln(2J$_0$+1) ≅ 0.69 the ground state pairing is accounted for by setting n to 0, 1 and 2 for even, odd and odd-odd nuclei. To have a continuous transition at $E_x(t_c)$ and to comply with the rules for a BCS system [37] the interpolation uses the auxiliary variable $\phi$, by setting (1- $\phi^2$) = ($E_x+E_{bs}$)/($E_t+E_{bs}$) = ($S-S_0$)·$t$/($S_t$·$t$). As was shown previously [23, 37 eq. A3, A5] the parameters $t$ and $d(t)$ are uniquely defined by $\phi$ and the energy dependence of $\rho(E_x)$ is characterized by a nearly constant logarithmic derivative of $\rho(E_x)$, the inverse of the 'nuclear temperature' $T$. As was pointed out [3], $T$ is usually larger than the parameter $t$. The results obtained for $T$ and $D(S_n,½^+) = \rho^{-1}(S_n,½^+)$ by using Eqs. 2 and 3 are compared in Fig. 2 to the experimental data compiled in the database of RIPL-3 [23]; it is depicted for more than

100 nuclei with A>70 and ground state spin ½⁺. For the region below $E_x(t_c)$ calculated averages of $T$ are compared in Fig. 2a to corresponding values extracted by various authors [23, 43, 44] from information on nuclear level schemes; in view of the scatter in these the agreement is satisfactory away from $^{208}$Pb. Fig. 2b shows the average distance $D(S_n,½^+)$ of s-wave resonances seen with neutron capture in even target nuclei [45, 23] in comparison to our prediction for the level distances at $S_n$ including the effect of triaxiality. As these all have spin $J_0=½^+$, a comparison of these data is free from spin cut-off ambiguities and it is worthwhile noting that for spin ½ the small $J$ limit differs from a more complete approximation by a few % only. Vibrational enhancement was investigated by inserting $\hbar\omega_{vib}=E_2(2^+)$ and $E_3(2^+)$ in the respective expression [4] with the energies $E_x(2^+)$ taken from the CHFB calculations; it would contribute less than 20%. Further studies are needed to clarify uncertainties in $\delta W_0$ [23] and in the influence of pairing on $\rho(E_x)$ below $E_t$, where a reduction was predicted for even nuclei [17], which is stronger than the one proposed elsewhere [37]. Global experimental data on the influence of parity on the level density is missing such that modifications may still result from the inclusion of parity effects, as well as from changes in the shell correction.

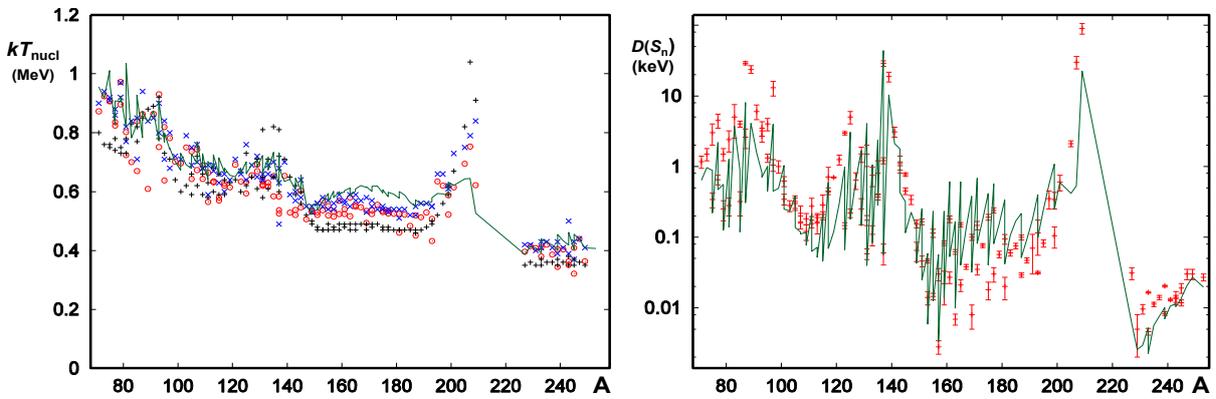

**Fig. 2:** Comparison of experimental level density information to predicted results presented as green line. **(a)**: Spectral temperature averaged between 1 MeV and $S_n$ **(b)**: Average resonance distance $D(S_n,½^+)$ vs. A. in comparison to experimental information (o [23], ✶[43], +[44]).

As seen from Fig. 2b nearly all of the measured resonance distances lie close to the prediction based on Eqs. (2) – (6). As already seen for the intrinsic level density $\omega(E_x)$ an important influence on $\rho(E_x,J)$ was found to emerge from the choice made for $\delta W_0$: Replacing the shell effect from ref. [40] by one of the others also listed [23] modifies the level density for actinide nuclei by up to a factor of two. As this difference is less for smaller A the A-dependence of $\delta W_0$ needs further theoretical study. For A ≈ 208 no agreement can be expected and it is of interest to study in detail, what reduction of collective enhancement near closed shells leads to an even better global fit.

## 4    Radiative neutron capture

The good agreement of the low energy slopes of the IVGDR to a 'triple Lorentzian' parameterization (TLO) as obtained by using independent information on triaxial nuclear deformation suggests the use of a corresponding photon strength function also for the radiative neutron capture, an electromagnetic processes alike. To test the influence of dipole strength functions on radiative neutron capture over a wide range in A the investigation of only even-even target nuclei has the advantage of offering a large sample with the same spin. For the $\ell$-wave capture by spin 0 nuclei the assumption $\Gamma_\gamma \ll \Gamma_n$ and the neglect of any $\ell$-dependent neutron strength enhancement leads to the cross section [46] :

$$\langle \sigma_R(n,\gamma) \rangle \cong 2(2\ell+1)\pi^2 \lambdabar_n^2 \cdot \rho(E_R,½^+)\cdot\langle\Gamma_{R\gamma}(E_\gamma)\rangle; \quad \langle\Gamma_{R\gamma}(E_\gamma)\rangle = \int M_t \rho(E_f,J_f) E_\gamma^3 f_1(E_\gamma) dE_\gamma. \quad (7)$$

The factor $M_t$ accounts for the number of magnetic sub-states reached by the γ-decay in comparison to the number of those populated by capturing the neutron. In view of Eq. (2) it is assumed here that for

$\lambda=1$-transitions from $J_R=1/2^+$ to $J_f=1/2$ and $J_f=3/2$ the quantum-statistical part of $M_t$ is 5. In the region well above separated resonances Porter-Thomas fluctuations [14, 15], albeit reduced by averaging over a large number of neutron resonances $R$, have to be corrected for. From statistical calculations a value of 0.87 was derived bringing $M_t$ to 4.4. It was pointed out previously [26] that strength information can be extracted from capture data directly by regarding average radiative widths $\langle\Gamma_\gamma\rangle$. Equation (7) shows, that these are proportional to the photon strength, and depend in addition on the ratio between the level densities at the capturing resonances - included in $f_1(E_\gamma)$ - and the final states reached by the γ-decay. Consequently the average radiative widths vary with the slope of $\rho(E_x)$ in the energy range reaching from $E_f$ to $E_R$, equivalent to the spectral temperature $T$ [3, 23, 43, 44], whereas capture cross sections also vary with the level density at $S_n$. A good agreement is found [18] between the $\langle\Gamma_\gamma\rangle$ predicted from TLO and average radiative widths as derived by a resonance analysis of neutron data taken just above $S_n$ and tabulated [45] for over 100 even-odd nuclei with A > 70.

As shown in Fig. 3a the agreement between predicted neutron capture cross sections for Th-, U- and Pu-nuclei and data is satisfactory on an absolute scale. As depicted for $^{238}$U the minor photon strength as discussed in the end of Ch. 2 is important: The dashed curve corresponds to TLO alone and the drawn one has the orbital M1 strength and E1 from $2^+\otimes3^-$ (cf. Ch.2) included as well. The electric dipole (pygmy) components other than isovector E1 known [24, 26-30, 35, 36] for higher $E_\gamma$ are suppressed by the steep decrease of $\omega(E_x)$ and strength at low $E_\gamma$ suffers from the factor $E_\gamma^3$ in Eq. (7) [28, 30]. The good agreement to actinide data within ≈ 30 % as seen in Fig. 3a (and for many other nuclei besides the ones shown) gives a convincing impression for the validity of the parameterization presented and the approximations applied.

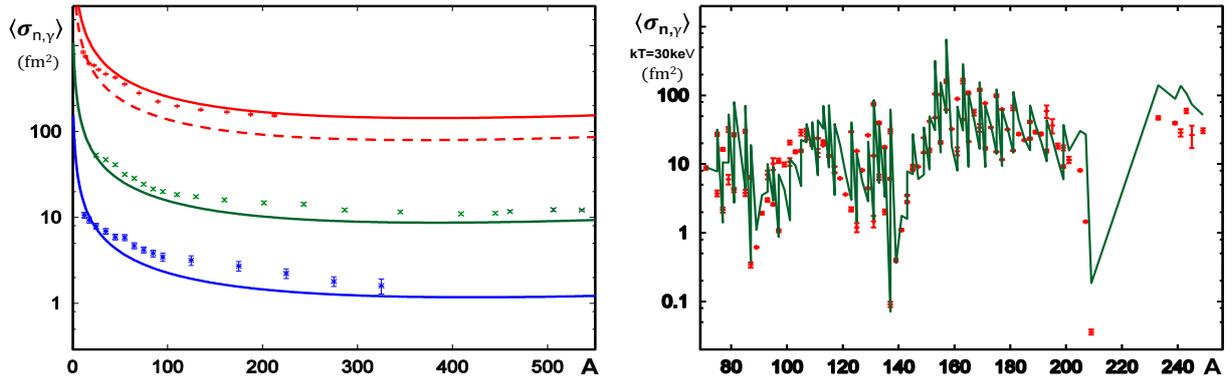

**Fig. 3:** Comparison of calculated neutron capture cross sections σ(n,γ) to experimental data (in fm²) [34].
**(a):** Dependence on $E_n$ for targets of (bottom to top) $^{240}$Pu (blue, ÷10), $^{238}$U (green) and $^{232}$Th (red, ×10).
**(b):** Maxwellian averaged cross sections vs. A for $kT_{AGB}$ = 30 keV.

To cover the full range of A>70 in the comparison to data Maxwellian averaged (MACS) neutron capture cross sections are shown in Fig. 3b together with the prediction made by folding of the cross sections as given by Eq. (7) with a Maxwellian distribution of neutron energies [2]. MACS have been tabulated [47] covering many heavy nuclei as they are of use for the investigation of nuclear processes in cosmic objects like red giant (AGB) stars, where radiative neutron capture takes place at approximately $kT_{AGB}$ = 30 keV. For several actinide nuclei equivalent data were compiled [48] and uncertainty bars were derived from the scatter as published. In view of the fact that $D \gg \Gamma_R \geq \Gamma_{R\gamma}$ the Maxwellian averages around 30 keV are formed incoherently and fluctuations (beyond the ones mentioned above) are neglected. By only regarding the radiative capture by spin-zero targets effects related to ambiguities of spin cut off parameter and angular momentum coupling are suppressed, but still the data vary by about 4 orders of magnitude in the discussed range of A – and they are well represented by the TLO-parameterization used here together with the schematic ansatz for $\rho(A, E_x)$, as described by Eqs. (2) – (6). The discrepancy observed in the region of A ≈ 90 may well be related to the present omission of p-capture, which is known to be especially important in that mass range [23, 37]. This and other local effects have no significance on the stated importance of triaxiality in heavy nuclei – the main topic here.

## 5     Conclusions

In agreement to various spectroscopic information available for a number of heavy nuclei with A > 70 [6, 9-12] two effects – hitherto not emphasised as such – indicate triaxiality for nearly all of them:

1) With one global parameter – and not five as usual [23] – the scheme proposed here reproduces observations for level densities in nuclei with $J_0 = ½$, when the collective enhancement due to symmetry reduction by triaxiality is included and the condensation energy $E_{con}$ is used for the Fermi gas backshift; this also avoids a recently detected [39] inconsistency.
2) Again only one global parameter suffices to fit to the shape of the IVGDR peak by a triple Lorentzian photon strength (TLO) – considerably improved and in accord to the TRK sum rule. It also predicts its low energy tail – without additional modification – to match respective strength data as well as neutron capture cross sections taken in the energy range of unresolved resonances.

For the last-mentioned finding a combination of the points 1) and 2) is needed, which is easily performed by considering spherical and axial symmetry to be broken – as shown by HFB calculations [8, 13]. Exact deformation parameters are unimportant for the tail of the E1-resonance as well as for the density of low spin states occurring in neutron capture by even targets as neither spin cut off nor moments of inertia are involved (cf. Eq. (2)). In addition to previous knowledge the triaxiality of most heavy nuclei is established here: For more than 100 spin-0 target nuclei with A>70 level distance data and average capture cross sections are well predicted by a global ansatz. The literature study performed within this work indicates a non-negligible effect of 'minor' magnetic and electric dipole strength other than isovector electric. Experimental photon data indicate that such strength may increase the radiative capture cross section by up to 60% and new experimental investigations of photon strength in the region of $E_\gamma$=3-5 MeV are desired. The global parameterization proposed here for isovector strength (TLO) with the discussed additions agrees well to radiative neutron capture cross sections [34] as shown in Fig. 3. It also does not exceed directly measured photon strength in the region below $S_n$ [5, 18, 19, 21, 24]. It can be considered a good ingredient for network calculations in the field of cosmic element production as well as for simulations of nuclear power systems and the transmutation of radioactive waste, were the applicability to actinide nuclei is of importance.

Previous work in the field of photon strength [*e.g.* 26] has worked with a lower IVGDR tail leading to a larger relative influence of 'minor' strength components. Here the often assumed dependence of the resonance widths on gamma-energy plays an important role. This is especially so if theory-based modifications [26] are added to seemingly improve $f_{E1}$ at small energies without much of a change in the peak region. Corresponding single or 2-pole IVGDR fits are likely to result in erroneous estimates of the corresponding E1-strength as they result in an irregular A-dependence of the spreading width $\Gamma_{E1}$ and the resonant cross section integral. This sheds some doubt on E1 strength predictions presented by RIPL [23] which obviously lead to such irregularities. In contrast the triple Lorentzian scheme (TLO) with a variation of $\Gamma_{E1}$ with the pole energy $E_0$ alone uses only one global parameter (the proportionality between $\Gamma_{E1}$ and $E_0$) and accords to the TRK sum rule resulting in a global dipole strength prediction for the tail region. The ansatz presented here assumes (at least weakly) triaxial shapes for nearly all heavy nuclei away from $^{208}$Pb. This finding is confirmed as the resulting collective enhancement improves the agreement to level distance data as well as to radiative capture cross sections. And the new level density description with only one global parameter results in a remarkable predictive power for compound nuclear reaction rates. Regarding the rather limited theoretical work done so far [6, 8, 13] the importance of broken axial symmetry already at low spin – as advocated here – should induce further investigations.

**Acknowledgements**

Discussions with K.H. Schmidt and R. Schwengner are gratefully acknowledged. This work was presented at the ERINDA workshop held at CERN in 2013 with support from EU-Fission-2010-4.2.1.